\documentclass[11pt,a4paper]{article}

\usepackage[utf8]{inputenc}
\usepackage[T1]{fontenc}
\usepackage{lmodern}
\usepackage{amsmath,amssymb,amsfonts}
\usepackage{graphicx}
\usepackage{hyperref}
\usepackage{geometry}
\usepackage{setspace}
\usepackage{enumitem}
\usepackage{url}
\usepackage{booktabs}
\hypersetup{colorlinks=true, linkcolor=blue, citecolor=blue, urlcolor=blue}

\begin{document}

% --- Custom title block to match original Word EXACTLY ---
\begin{center}
{\Large\bfseries The tragedy of the cognitive commons: collective intelligence beyond AI-induced knowledge collapse\par}
\vspace{1.5em}

{\normalsize Maher Kallel\par}
{\small Senior Strategic Consultant\par}
{\small Email: maher.kallel@gmail.com\par}
{\small ORCID: 0009-0006-2615-4392\par}
\vspace{1.2em}
{\small and\par}
\vspace{1.2em}
{\normalsize Mohamed El Louadi\par}
{\small Institut Sup\'erieur de Gestion, Universit\'e de Tunis\par}
{\small 41 rue de la Libert\'e -- Cit\'e Bouchoucha, 2000 Le Bardo, Tunisia\par}
{\small Email: mohamed.louadi@isg.rnu.tn\par}
{\small ORCID: 0000-0003-1321-4967\par}
\end{center}
\vspace{2em}

\begin{abstract}

A recent dynamic model by Acemoglu, Kong, and Ozdaglar (2026) argues that agentic AI can trigger a self-reinforcing erosion of humanity's shared knowledge base, a phenomenon the authors term knowledge collapse. The model rests on a complementarity between slowly accumulated general knowledge and locally generated context-specific knowledge, and on a learning externality whereby individual effort jointly produces a private signal and a thin public signal that feeds the collective stock. When agentic AI substitutes for the private signal without replenishing the public one, and when human effort is sufficiently elastic, the system can converge on a low-knowledge equilibrium in which welfare is non-monotonic in AI accuracy. This paper offers a measured appraisal. It distinguishes what the model demonstrates from what popular summaries claim, situates it within the literature on knowledge commons and model collapse, and weighs the partial empirical evidence, including a documented 25\% decline in public knowledge sharing on Stack Overflow. It then advances five structural critiques, targeting the model's fixed knowledge taxonomy, its assumption that AI cannot produce general knowledge, its pivotal unmeasured parameter of effort elasticity, the long historical record of comparable cognitive alarms, and the political infeasibility of its proposed remedy. The paper concludes that the model's value lies in identifying the tragedy of the cognitive commons, a credible negative externality. Catastrophe is not guaranteed, and the remedy is stronger aggregation of validated human knowledge. The paper sets out a research agenda for measuring effort elasticity and monitoring the commons.
\end{abstract}


\section{Introduction}

Few theoretical artifacts travel as quickly, or as carelessly, as a pessimistic model of artificial intelligence authored by a Nobel laureate. Within days of its circulation as National Bureau of Economic Research Working Paper 34910, the analysis of Acemoglu, Kong, and Ozdaglar (2026) had been compressed on social platforms into a sequence of arresting claims: that artificial intelligence will mathematically guarantee the destruction of human knowledge, that greater machine precision is strictly worse for society, and that a Nobel Prize confers the authority of proof on the conclusion. None of these statements survives contact with the paper itself. Yet the underlying contribution is serious, and it deserves neither the hysterical amplification it has received nor the reflexive dismissal that such amplification tends to provoke.

This paper offers a critical perspective whose purpose is to provide that intermediate appraisal for a readership concerned with technological forecasting and the governance of socio-technical transitions. The exercise matters because the model articulates, in formal terms, a class of risk that conventional productivity-centered assessments of AI tend to ignore. Most macroeconomic treatments, including Acemoglu's own earlier work on the aggregate effects of automation (Acemoglu, 2024; Acemoglu \& Restrepo, 2019), evaluate AI through the static lens of task displacement and reinstatement. The knowledge-collapse model instead foregrounds a dynamic externality operating on the stock of collective understanding, an effect that is invisible to a snapshot of productivity gains and that may only become legible over a horizon of years or decades. Forecasting institutions that miss this channel risk systematically underestimating the long-run costs of cognitive automation, even as they correctly measure its short-run benefits. The paper has a dual aim: It is in part a critique, correcting both the model's assumptions and its viral misreading, and in part a constructive contribution, reframing the phenomenon as a tragedy of the cognitive commons and proposing the measurement and institutional agenda that this reframing implies.

The argument develops in several steps. Section 2 reconstructs the model's logic, taking care to state precisely what it asserts and under what conditions, and closes with a formal sketch that isolates the parameters on which the result depends. Section 3 surveys the convergent empirical signals that lend the mechanism partial plausibility while stopping well short of confirming it at societal scale, and weighs the countervailing evidence on effort reallocation. Section 4 develops five structural critiques. Section 5 sets out what calibration and falsification of the model would require, including a design proposal for measuring its pivotal parameter. Section 6 separates the model's defensible content from the rhetorical slippages that have accompanied its popular diffusion. Section 7 identifies what is genuinely durable in the contribution and translates it into a forward-looking research and policy agenda. A short conclusion follows.

\section{The model and what it actually claims}

The model's architecture can be decomposed into three interlocking components: an epistemic dichotomy between two kinds of knowledge, a feedback mechanism that links them through human effort, and a welfare result that follows from the first two. Each is reconstructed in turn, after which a formal sketch isolates the parameters that govern the outcome.

\subsection{An epistemic dichotomy}

The model is built upon a deliberately sharp distinction between two kinds of knowledge, general knowledge and context-specific knowledge. General knowledge is a social and collective stock, accumulated slowly through shared learning: the principles of pathology that organize clinical reasoning, the mechanisms of price formation that structure financial judgment, the accumulated doctrine of a legal tradition, etc. Context-specific knowledge is local and private, attached to a particular situation: the constellation of symptoms in front of a given clinician, the composition of a single investor's portfolio, the facts of one client's case.

The decisive modeling assumption is that these two forms are complements, with limited potential for substitution (El Louadi, 2024). A context-specific recommendation, however precise, is informative only to an agent who possesses the general framework needed to interpret it. A perfectly accurate signal about a particular case becomes noise if no one retains the conceptual scaffolding within which it acquires meaning. This complementarity is the structural pivot on which the entire argument turns, and it is, as Section 4 argues, both the model's principal source of insight and one of its more contestable premises.

\subsection{The mechanics of collapse}

In the model, human cognitive effort produces both kinds of knowledge simultaneously through an economy of scope in learning. When a physician works through a difficult diagnosis, the effort generates a private signal about the specific patient and, as a by-product, a thin public signal that diffuses into the collective stock of medical understanding. This joint production is the engine of a learning externality: the individual bears the full cost of effort but captures only the private return, while the public return accrues diffusely to the community. The social value of effort therefore exceeds its private value, and the market under-provides it even before AI enters the picture.

In our framing, agentic AI substitutes for the private signal without ever generating the public one. It answers the immediate question while contributing nothing to the shared stock. From this asymmetry a negative feedback loop emerges. Accurate machine recommendations lower the private return to human effort; reduced effort thins the flow of public signals; the stock of general knowledge degrades; and the degraded stock makes the AI's own recommendations less reliable, because the machine ultimately draws on the same collective substrate. The system consumes the very resource that made it valuable. The companion analysis of how aggregation mechanisms shape this dynamic (Acemoglu, Lin, et al., 2026) reinforces the centrality of the public-signal channel.

The crucial point, frequently lost in summary, is that collapse is a conditional outcome. It emerges only when human effort is sufficiently elastic, meaning that agents sharply reduce their effort once AI satisfies their immediate needs. When effort remains inelastic, sustained by professional norms, intrinsic curiosity, or institutional obligation, the public signal persists and the catastrophic equilibrium remains unlikely. The model identifies a possibility that depends on specific conditions; its results should therefore be interpreted as a conditional theoretical mechanism.

\subsection{The non-monotonicity of welfare}

The model's most counter-intuitive result is about the relationship between AI accuracy and social welfare. Welfare is not monotonically increasing in machine precision. At low levels of accuracy, AI improves decisions and allows for net positive gains. Beyond a critical threshold, the dynamic losses from eroded general knowledge come to dominate the static gains from better-personalized recommendations. The implication is that there may exist a socially optimal level of AI that is deliberately imprecise, and that information-design regulation might be justified to hold accuracy below the threshold.

This is the result that popular accounts have rendered as more precise AI is always worse. The model says something narrower and more interesting: the marginal social value of precision can turn negative past a point, conditional on the elasticity of effort and the complementarity assumption. A claim about a threshold and a conditional sign reversal is a hypothesis amenable to calibration and falsification; a claim of monotone immiseration is neither.

\subsection{A formal sketch of the feedback loop}

To fix intuition and to expose the quantities that policy and measurement must target, it is useful to reconstruct the mechanism in a stylized form. What follows is not a restatement of the authors' full apparatus but a heuristic reduction in its spirit, designed to make the threshold condition and the welfare non-monotonicity legible to a forecasting readership and to identify the parameters that a calibration exercise would have to pin down. Figure 1 summarizes the resulting feedback structure.

\begin{figure}[htbp]
\centering
\includegraphics[width=0.96\textwidth]{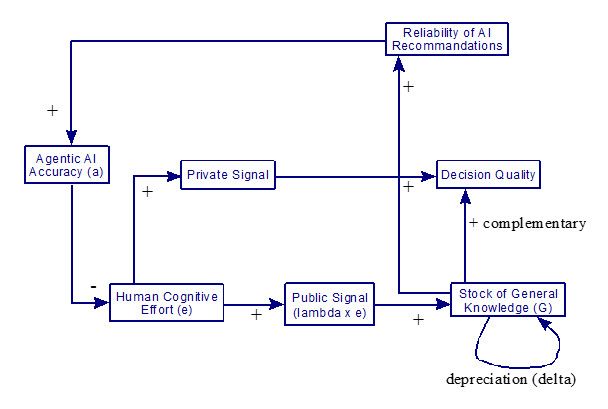}
\caption{Causal structure of the knowledge-collapse mechanism. Human cognitive effort jointly produces a private signal, which raises immediate decision quality, and a public signal, which replenishes the stock of general knowledge $G$. General knowledge complements the private signal in decision quality and sustains the reliability of AI recommendations. Agentic AI accuracy substitutes for the private signal and so reduces the marginal return to effort (the negative link shown in red), thinning the public signal and eroding $G$, which in turn degrades the recommendations that depend on it. The self-undermining loop runs from effort to the public signal to $G$ to AI reliability and back; $\delta$ denotes depreciation of general knowledge and $\lambda$ the productivity of the public signal.}
\label{fig:knowledge-collapse}
\end{figure}

Let $G \in [0,1]$ denote the stock of general knowledge in the relevant domain. A representative agent chooses cognitive effort $e$ at convex cost $c(e)$, with $c'(e)>0$ and $c''(e)>0$. Effort yields two products jointly: a private signal about the agent's own context, which improves the immediate decision, and a thin public contribution to the common stock, taken to be proportional to effort at rate $\lambda$. Agentic AI supplies, at negligible marginal cost, a recommendation of accuracy $a \in [0,1]$ that substitutes for the private signal alone. The parameters of the sketch are as follows:

$G$: the stock of general knowledge, or commons, normalized to $[0,1]$;

$e$ and $c(e)$: individual cognitive effort and its convex cost;

$a \in [0,1]$: the accuracy of the agentic AI recommendation;

$\lambda$: the public-signal productivity, the rate at which effort replenishes $G$;

$\delta$: the depreciation rate of general knowledge;

$\varepsilon$: the elasticity of effort with respect to AI accuracy;

$D$: the social shadow value of general knowledge.

Decision quality combines the two knowledge forms through a complementarity. Write it as $Q(G,s)$, where $s$ is the available context-specific signal, with $\frac{\partial^2 Q}{\partial G \partial s}>0$: the marginal value of a context-specific signal rises with the stock of general knowledge, and conversely. This single inequality is the formal heart of the model. It implies that a context signal delivered by AI, however accurate, contributes little to decision quality when $G$ is low, which is what makes the erosion of $G$ self-limiting for the machine that depends on it.

The agent's privately optimal effort solves the first-order condition that the marginal private benefit of effort equals its marginal cost. Because AI substitutes for the private signal, the marginal private benefit of effort is decreasing in accuracy $a$: when the machine already supplies a good context signal, the agent's own effort buys less incremental decision quality. Optimal effort is therefore a function $e^*(a,G)$, decreasing in $a$. The key behavioral primitive is the elasticity of effort with respect to accuracy,
\begin{equation}
\varepsilon = -\frac{\partial e^*}{\partial a}\cdot\frac{a}{e^*},
\tag{1}
\end{equation}
which measures how sharply agents withdraw effort as the machine improves. This elasticity is not estimated in the source paper and is, as argued below, the pivotal unknown.

The common stock evolves according to a law of motion in which it depreciates at rate $\delta$ and is replenished by the aggregate public signal,
\begin{equation}
G' = (1-\delta) G + \lambda\, \mathbb{E} e^*(a,G)
\tag{2}
\end{equation}
where $\mathbb{E}$ denotes the average over agents and $G'$ is next period's stock. A steady state $G^*(a)$ satisfies
\begin{equation}
\delta\, G^*(a) = \lambda\, \mathbb{E}\, e^*(a, G^*(a))
\tag{3}
\end{equation}

Because $e^*$ is decreasing in $a$, the steady-state stock $G^*(a)$ is also decreasing in $a$: more accurate AI sustains a thinner commons. When the complementarity is strong and effort is sufficiently elastic, the steady-state map can become non-monotone, taking an S-shape that admits multiple fixed points: a knowledge-rich equilibrium and a knowledge-poor one separated by an unstable threshold. Collapse is then the loss of stability of the knowledge-rich equilibrium as accuracy rises, so that trajectories beginning below the threshold are drawn to the low-knowledge fixed point. This is the formal content of collapse: a transition marked by the disappearance of the high-knowledge basin of attraction once $a$ crosses a critical value $a^\star$ which depends positively on the depreciation rate $\delta$ and the effort elasticity $\varepsilon$ and negatively on the public-signal productivity $\lambda$. The qualitative claim, that multiplicity and a tipping threshold arise under strong complementarity and high elasticity, is what this heuristic is meant to convey; a full treatment would derive the existence, uniqueness, and stability conditions formally, which this reduction does not attempt and which the original paper supplies.

Social welfare adds to private decision quality the value of the externality that agents ignore. A reduced-form expression captures the tension,
\begin{equation}
W(a) = S(a) - D\,[G^{(0)} - G^*(a)]
\tag{4}
\end{equation}
where $S(a)$ is the static surplus from better personalized recommendations, increasing and concave in $a$, and the second term is the dynamic loss from a degraded commons, with $D$ the social shadow value of general knowledge and $G^{(0)}$ the knowledge-rich benchmark. Differentiating, welfare rises with accuracy while the marginal static gain $S'(a)$ exceeds the marginal dynamic loss $D\,|\partial G^*/\partial a|$, and falls once the inequality reverses. The crossing point defines the socially optimal accuracy $a^{\text{opt}}$, which is interior and, under mild conditions on the shadow value $D$, strictly below the technological maximum whenever the dynamic term is active. The popular rendering that more precision is always worse corresponds to the special and implausible case in which the dynamic term dominates everywhere; the model's actual claim is the interior optimum, which is both milder and more useful.

Three features of this sketch deserve emphasis because they structure everything that follows. First, the sign of the policy-relevant comparative static depends on three quantities: the effort elasticity $\varepsilon$, the depreciation rate $\delta$ of general knowledge, and the public-signal productivity $\lambda$. Second, the pessimism depends on parameter values and assumptions embedded in the model: for low $\varepsilon$, high $\lambda$, or low $\delta$, the interior optimum sits at or near the technological frontier and there is no meaningful collapse region. Third, two simplifications should be made explicit, because the critique developed below turns on them. The representative-agent formulation suppresses the heterogeneity in effort elasticity that Section 4.3 argues is decisive; a fuller treatment would replace the scalar $\varepsilon$ with a distribution across domains, and the aggregate outcome would then depend on the shape of that distribution. And the shadow value $D$ is treated as constant, whereas it is plausibly endogenous to $G$ through the same complementarity, which would steepen the dynamic loss term near the threshold and push the socially optimal accuracy $a^{\text{opt}}$ further below the technological frontier. With those caveats registered, the elasticity $\varepsilon$ and the depreciation rate $\delta$ are estimable from field data, while the public-signal productivity $\lambda$ is the hardest to identify and would require strong exclusion restrictions. This is precisely why the appropriate response to the model is measurement, free from alarm or dismissal.

\section{The empirical signals}

The mechanism is not purely speculative. Several independent empirical findings give it partial support, though none establishes collapse at the level of society.

The most directly relevant evidence concerns public knowledge sharing. Studying activity on Stack Overflow, the principal public repository of programming knowledge, del Rio-Chanona et al.~(2024) document a decline of roughly one quarter in question volume in the six months following the release of ChatGPT (November 30, 2022), relative to comparable platforms and to regions with restricted access. The decline has continued, returning activity to levels last seen years earlier. The interpretation matters: the platform's content is not being shown to be wrong, and the underlying problems are not disappearing. What is disappearing is the act of reasoning in public. Private resolution through a conversational assistant displaces the externality-generating behavior on which the commons depended. This is consistent with the thinning of the public signal that the model posits, observed in a single high-resolution domain, although migration of the same reasoning to less visible venues cannot be excluded from this design alone, a point developed below.

A second strand concerns the individual cognitive substrate. In an experimental study combining electroencephalography with linguistic analysis, Kosmyna et al.~(2025) find that participants writing essays with a large language model exhibited lower measured neural connectivity than participants writing unaided, and that a substantial majority were subsequently unable to quote from essays they had nominally authored. When later asked to write without assistance, prior AI users underperformed those who had never used the tool. The authors describe this pattern as the accumulation of cognitive debt. The finding speaks to the demand side of the model, specifically the willingness and capacity to expend cognitive effort, although it concerns individual cognition while leaving the collective stock unexamined. Its evidential weight should not be overstated: it is a small, short-horizon laboratory study on a scaffolded essay task, the inference from reduced electroencephalographic connectivity to cognitive impairment is contested among neuroscientists, and the persistence and ecological validity of the effect are unestablished. It provides, at most, weak evidence for the demand-side channel, with its broader validity remaining unestablished.

A third, more distant analogy comes from machine learning itself. Shumailov et al.~(2024) demonstrate that generative models trained recursively on their own outputs degenerate, a phenomenon they call model collapse, in which the tails of the data distribution are progressively lost. The parallel provides a useful analogy, although it does not constitute direct evidence: where model collapse describes degradation in the machine's training substrate, the Acemoglu model describes degradation in the human knowledge substrate, and both identify a failure mode in which a self-referential learning system erodes its own informational foundation. The disanalogy is nonetheless fundamental because the two mechanisms differ in kind: model collapse is a statistical property of recursive training on truncated distributions, whereas the human mechanism is mediated by incentives and institutions, so the parallel motivates the question without supplying evidence for the answer.

A fourth body of evidence cuts the other way and must be weighed carefully. Field studies of generative AI in the workplace document substantial productivity gains, with effort being redirected toward other activities. Customer-support agents assisted by a large language model resolved issues faster and the gains concentrated among less-experienced workers (Brynjolfsson et al., 2025), while professionals assigned a generative tool completed writing tasks more quickly and at higher quality (Noy \& Zhang, 2023). These findings do not contradict the collapse mechanism, since they capture private productivity while leaving effects on the public stock unresolved; however, they caution against assuming that effort released from one task simply evaporates. If freed effort is redeployed to higher-order problems that themselves generate public signals, the net effect on the commons is ambiguous and possibly positive. The empirical record, in other words, identifies the demand-side withdrawal the model needs but also the reallocation channel the model omits. This caveat carries the weight of a potential falsifier: were systematic redeployment of freed effort toward public-signal-generating activity to be established, the central mechanism would fail outright, which is why Section 5.2 treats reallocation as a central empirical test.

These signals also raise a hard identification problem that the forecasting community should not gloss. The Stack Overflow decline is consistent with public-signal thinning, but it is equally consistent with a migration of the same reasoning to less visible venues, private chats, internal wikis, model-training corpora, in which case the commons is not shrinking but relocating beyond the reach of the instrument. Distinguishing genuine erosion from measurement displacement is the central empirical challenge, and it cannot be resolved by any single platform metric. The cognitive-debt findings, for their part, are short-horizon laboratory results whose persistence and external validity remain open. None of this dissolves the concern; it locates it. The mechanism is real where it has been observed, its aggregate sign is undetermined, and the determination requires designs built specifically to separate erosion from relocation and substitution from reallocation.

Taken together these signals are convergent and informative. They are also partial. Each illuminates one link in the posited chain, public-signal thinning, effort withdrawal, and substrate degradation in a bounded setting. None observes the full feedback loop closing at societal scale, and none rules out the compensating mechanisms examined below. The honest summary is that the mechanism is plausible and locally evidenced, with societal-scale collapse remaining undocumented.

\section{A critical appraisal}

Five structural objections emerge from a close reading. They bear, in turn, on the fixity of the knowledge taxonomy, the assumption that AI cannot produce general knowledge, the unmeasured parameter on which the result depends, the long historical record of comparable alarms, and the feasibility of the proposed remedy. A final subsection turns the same critical lens on this appraisal itself.

\subsection{A frozen taxonomy of knowledge}

The most fundamental structural objection concerns the fixity of the model's central dichotomy. The framework captures the erosion of existing categories of knowledge but cannot represent the emergence of new categories that AI itself calls into being. The partition into general and context-specific knowledge is held constant, yet the history of cognitive technologies is in large part a history of taxonomic change, of competences that did not previously exist becoming central. Early commentary on the paper has pressed exactly this point, noting that the model has no place for hybrid human-machine forms of intelligence that lack a settled name (Cordasco, 2026; Galiani, 2026).

This matters because the welfare accounting depends on the taxonomy. If AI dissolves certain forms of general knowledge while simultaneously generating new forms of distributed, human-machine competence, then a model that scores only the loss will overstate the net cost. The point is that the adequacy of these new competences as compensation remains an open question, and a static taxonomy cannot even pose it. The model's pessimism is, in part, an artifact of holding fixed the very thing that powerful general-purpose technologies tend to transform. A more adequate framework would classify knowledge not by fixed type but by its evolving role in the human-machine division of cognitive labor, allowing categories to be created and reclassified as the technology matures. Such a functional taxonomy would let the same formal machinery register the birth of new competences alongside the erosion of old ones, though specifying it operationally remains an open problem that this critique identifies without fully resolving. Which competences are counted as general knowledge worth protecting is, moreover, a normative question. Classification encodes whose expertise is valued, and a taxonomy frozen at today's professional categories risks naturalizing existing knowledge hierarchies instead of critically examining them, a concern the welfare accounting should register even though the model abstracts from it.

One constructive way to make the taxonomy dynamic is to classify general knowledge according to cognitive levels instead of traditional domains. Declarative knowledge (facts), procedural knowledge (skills), causal knowledge (understanding of why), and metacognitive knowledge (the capacity to learn and to judge) are affected by AI very differently. Agentic systems substitute most readily for declarative recall and routine procedure, less easily for causal understanding, and least of all for the metacognitive capacity to frame problems and to evaluate machine output. A framework that scored erosion at the declarative level while crediting possible gains at the metacognitive level would reach conclusions the single aggregate $G$ cannot represent, and would identify the domains in which human effort retains a complementary role alongside the machine.

\subsection{Can AI produce general knowledge?}

The framework assumes that agentic AI recycles general knowledge without producing it. This is a strong premise, and the evidence increasingly strains it. Scientific foundation models already generate outputs that are general knowledge by any reasonable definition. The protein-structure predictions of AlphaFold (Jumper et al., 2021) constitute additions to the shared stock of biological understanding: new structures, new hypotheses, new theoretical regularities available to the entire community. These constitute durable public contributions that remain available to the broader scientific community.

If AI contributes to the production of general knowledge alongside its use of existing knowledge, the feedback loop is no longer unidirectional. The boundary between recycling and creating is more porous than the model admits, and the substantial private and public investment now flowing into AI for scientific discovery, whatever the eventual return, reflects a working assumption that these systems can generate genuine knowledge. A complete model would treat AI's contribution to the general stock as an endogenous quantity that policy and design can raise, not as a parameter fixed at zero.

A sharper version of the same objection targets the public-signal productivity itself. The model fixes the rate $\lambda$ at which human effort replenishes the commons, yet AI plausibly raises it. An assistant that helps a clinician turn a difficult case into a publishable report, or a developer turn a one-off fix into reusable documentation, increases the public yield of a given unit of effort through knowledge sharing and reuse, while leaving the accuracy of private recommendations as a separate dimension; in the notation of Section 2.4, it acts on $\lambda$, not on $a$. If agentic AI lowers the cost of converting private insight into public contribution, it can thicken the commons even as it reduces the raw quantity of effort, so that the sign of its net effect becomes an empirical question about which margin dominates, awaiting empirical resolution.

Both objections can be absorbed into the formal sketch by extending the law of motion (2) with a term for the machine's own contribution,
\begin{equation}
G' = (1-\delta) G + \lambda\, \mathbb{E} e^*(a,G) + \gamma\, A(a)
\tag{5}
\end{equation}
where $\gamma$ is the rate at which AI adds directly to the general stock and $A(a)$ its output at accuracy $a$. With $\gamma>0$, or with AI raising $\lambda$ as argued above, the steady-state stock need not decline with $a$: when the machine's contribution and its enhancement of public-signal productivity together outweigh the effort it displaces, $G^*(a)$ can increase and the collapse region can shrink or disappear. Systems that already produce durable general knowledge, AlphaFold in structural biology and comparable models in mathematics and weather prediction, indicate that $\gamma$ is not zero, which makes estimating it as urgent as estimating the effort elasticity. The model's most consequential assumption, $\gamma=0$, is thus also its most empirically vulnerable.

\subsection{Effort elasticity: the unmeasured linchpin}

The model is explicitly conditional, and the condition that does the work is the elasticity of human effort. Collapse requires that individuals withdraw cognitive effort sharply when AI satisfies their immediate need. Everything turns on this parameter, and the paper does not measure it.

There are strong reasons to expect heterogeneity. Effort elasticity varies with professional norms, occupational incentives, intrinsic motivation, and institutional obligations. Physicians, lawyers, and engineers do not abandon their training because a machine answers well; their effort is sustained by liability, licensure, reputation, and the internal goods of practice. In other domains, where effort is purely instrumental and externally unmonitored, elasticity may be high. The aggregate dynamics depend on a distribution of elasticities across domains that the framework reduces to a single representative parameter. Whether the economy as a whole sits in the collapse region is therefore an empirical question that the theory frames but cannot answer. This is the single most important target for future measurement, and Section 5 returns to it with a concrete design.

\subsection{The long history of cognitive alarm}

Every cognitively powerful technology has caused the same fear. In the Phaedrus, Socrates warns that writing will implant forgetfulness in the soul, that learners will cease to exercise memory and will mistake the appearance of wisdom for its substance (Plato, ca. 370 BCE/1995). The printing press, the telephone, and the internet each attracted structurally identical anxieties. The historical record is sobering for the alarmist position. Writing did not hollow out memory so much as externalize and extend it; printing did not erode general knowledge but massified it; the calculator did not destroy mathematicians but raised the level of what is demanded of them.

This precedent counsels caution. The recurrence of the alarm does not prove that the present case is another false positive, since the structural features of agentic AI may differ in kind from earlier tools. The relevant distinction is one the distributed-cognition tradition has long emphasized. Earlier cognitive technologies largely externalized and extended cognition, redistributing it across people and artifacts while leaving the human reasoning process intact and often enriching it (Hutchins, 1995; Clark \& Chalmers, 1998). Agentic AI is the first to offer a substitute for the inferential step itself, not merely for the storage or transmission of its products, and it is on this difference, independent of the bare fact of novelty, that any claim of discontinuity must rest. But the base rate is informative. A long sequence of confident predictions of cognitive decline, each subsequently falsified by adaptive institutional and individual responses, should lower the prior assigned to any new instance, including this one. It is worth recalling, too, that critiques of Acemoglu's broader stance on AI have often argued that his estimates understate AI's potential effects, which complicates any interpretation of the present model as a settled forecast.

\subsection{The political infeasibility of the remedy}

The model's normative recommendation is to limit AI precision deliberately or, alternatively, to strengthen the aggregation of human knowledge. The first horn faces an obstacle the paper underweights. No private developer has any incentive to degrade their own product, and any unilateral restraint is competitively self-defeating. Coordinated restraint would require international regulation of model accuracy, a governance problem of great difficulty, both technically (because accuracy is multidimensional and hard to specify), and politically (because verification across jurisdictions is weak). Two qualifications soften the verdict without overturning it. Precision can be governed sectorally even where it cannot be governed globally, since several professions already constrain the form and disclosure of information-based advice, and voluntary norms or mandated usage warnings, on the model of pharmaceutical labeling, offer a lighter instrument than a hard accuracy cap. These are partial and domain-bounded remedies that leave the international coordination problem unresolved, yet they suggest that implementation constraints vary across contexts.

The difficulty is compounded by a geopolitical tension. The reforms most necessary to preserve collective cognition are precisely those that threaten the informational rents on which strategic advantage in AI is built, an asymmetry that is especially acute for economies specialized in information intermediation (Fetzer, 2026). A state that leads in model capability has limited interest in a regime that caps capability, and the distribution of gains from any such regime is sharply contested. The recommendation to throttle precision is therefore difficult to implement because it conflicts with the incentive structure that governs the technology's development. This asymmetry is one reason the second horn of the recommendation, strengthening aggregation, is far more promising, as Section 7 argues.

\subsection{Limitations of this appraisal}

Symmetry requires turning these standards on the present analysis. Three limitations should be acknowledged. First, the evidence assembled in Section 3 reflects a purposive selection of studies and is not intended as a systematic review; a comprehensive search might shift the balance of signals in either direction. Second, the formal sketch of Section 2.4 is the present authors' heuristic reduction, designed to clarify the mechanism without reproducing a verified derivation of the original model; it may misstate features of an apparatus it deliberately simplifies. Its role is to aid intuition and to locate the measurable parameters, not to substitute for the source paper's proofs. Third, the object of critique is a working paper that its authors may revise before journal publication, so specific points may be overtaken even as the broader appraisal stands. None of these caveats weakens the central argument, which concerns the structure of the externality and the logic of the policy response, but each bounds the confidence appropriate to the particulars.

\section{Calibration and falsifiability}

A theoretical model earns its place in forecasting through the discipline of its testable content, which matters more than the drama of its conclusion. The formal sketch of Section 2.4 isolates three parameters that jointly determine whether an economy, or a particular domain within it, occupies the collapse region: the effort elasticity $\varepsilon$, the depreciation rate $\delta$ of general knowledge, and the public-signal productivity $\lambda$. Each is, at least in principle, estimable, and the model makes distinct predictions depending on their values. This section sets out what calibration would require and what observations would falsify the mechanism, because a hypothesis that cannot fail is not a contribution to forecasting.

\subsection{What the model predicts}

The framework yields several qualitative predictions that are sharper than the popular summary suggests. First, public-signal generation should decline in domains where AI accuracy is high and effort is driven primarily by instrumental incentives, while showing limited or no decline where professional, reputational, or licensing structures sustain engagement. The prediction is therefore one of heterogeneity across domains, and findings of uniform decline or uniform stability would both be awkward for the theory. Second, the decline should primarily affect the public component of effort, namely the sharing and externalization of reasoning, whereas private task completion may even improve, according to productivity studies. Third, and most distinctively, the model predicts an eventual degradation in the quality of AI recommendations themselves in domains that have passed the threshold, as the thinning commons feeds back into the systems that draw on it, the dynamic captured by the law of motion in equation (2). This last prediction is the model's most vulnerable and therefore its most valuable: it is the point at which the theory stakes a claim that the optimistic alternative, in which AI and human knowledge co-evolve productively, would not make.

\subsection{What would falsify the mechanism}

The mechanism would be substantially undermined by any of the following observations. If domains with high AI penetration show no decline in public-signal generation once relocation and substitution are properly accounted for, the demand-side premise fails. If effort released by AI is systematically redeployed to public-signal-generating activities of higher order, the net externality may be positive and the collapse framing misleading. If AI systems demonstrably add to the stock of general knowledge at a rate comparable to or exceeding the human contribution they displace, the unidirectional assumption of Section 4.2 collapses and with it the feedback loop. And if recommendation quality in high-penetration domains fails to degrade over a horizon long enough for the commons effect to operate, the model's most distinctive prediction is disconfirmed. None of these tests is trivial, but none is beyond the reach of careful empirical design, which is the relevant standard.

\subsection{Measuring effort elasticity: a design proposal}

Because effort elasticity is the pivotal unknown, it warrants a dedicated measurement strategy that directly targets this parameter. A credible design would exploit the staggered introduction of capable AI assistants across teams, firms, or jurisdictions, the same source of variation that has powered recent workplace studies of generative AI (Brynjolfsson et al., 2025; Noy \& Zhang, 2023), and would estimate the response of effort to accuracy using a difference-in-differences or event-study specification around each rollout.

The design hinges on measuring the right outcome. The quantity of interest is the public component of effort. Measuring it requires indicators such as the rate of contribution to shared repositories, the frequency and novelty of publicly externalized reasoning, the production of documentation and explanations beyond private resolution, and the diversity of problems addressed publicly. Treatment intensity can be proxied by the accuracy or adoption of the assistant in each unit, allowing the slope of effort with respect to accuracy, and hence the elasticity $\varepsilon$ of equation (1), to be recovered. The central inferential threat, already noted, is the displacement of reasoning to unobserved venues; addressing it requires either instrumenting the visible commons with private-channel data through partnership agreements or triangulating across multiple public venues so that relocation within the observed set can be distinguished from genuine withdrawal. Two refinements sharpen the design. The outcome must be defined operationally, whether as hours of unaided work, the count of decisions taken without consulting the assistant, or validated public contributions per period, since each captures a different facet of effort. And the response must be decomposed into an intensive margin, less effort per task, and an extensive margin, the abandonment of whole categories of public-signal-generating activity; the dynamics of collapse are driven mainly by the latter, which a design measuring only average effort per task would miss. Separating the two requires tracking both the intensity of contribution among those who remain active and the rate of exit from the contributor pool, since only the second isolates the extensive margin.

Two further cautions are essential, and they qualify the strength of any such estimate. First, a staggered-rollout design of this kind recovers, at best, the reduced-form elasticity of observable public contributions with respect to AI adoption, which is not the structural parameter $\varepsilon$ of the model. The structure additionally requires separating the private-to-public split of effort, for which adoption alone is not a valid instrument; conflating the two would mistake a contribution elasticity for the behavioral primitive. One candidate route is to exploit variation in professional licensing or credentialing that shifts the private-to-public split independently of AI adoption, supplying the exclusion restriction that adoption alone lacks. Although speculative, this approach suggests that the structural gap is reducible and may close over time. Second, the recent econometrics of staggered difference-in-differences shows that conventional two-way fixed-effects estimators are biased under heterogeneous and dynamic treatment effects (de Chaisemartin \& D'Haultfoeuille, 2020; Callaway \& Sant'Anna, 2021), so a credible design must use a heterogeneity-robust estimator. A related limitation is temporal: event studies identify the short-run response of effort, whereas the steady-state dynamics that drive collapse depend on the long-run elasticity, and the two can diverge substantially as norms and institutions adapt. Heterogeneity analysis (El Louadi, 2024) across occupations would then map the distribution of elasticities that the aggregate dynamics require, and would identify the specific domains, if any, that sit in the collapse region. Estimates of this distribution, interpreted with these caveats, would begin to convert the model from a parable into a calibrated instrument, and would tell policymakers not whether to worry in general but where to look first.

\section{Disentangling the model from its viral image}

The gap between the paper and its circulating summary is wide enough to warrant explicit correction, because the popular version is the version that will shape lay intuition and, potentially, policy. Several slippages recur. The claim that collapse is mathematically guaranteed misrepresents a conditional result that holds only under high effort elasticity and above a precision threshold; change either parameter and the same mathematics yields the benign coexistence of capable AI and a thriving commons. 

The claim that AI will destroy human knowledge conflates a possible erosion of the collective stock of general knowledge with the destruction of individual cognition, which the model does not assert. The claim that more precise AI is always worse flattens a non-monotonicity, in which gains are real and positive at lower precision, into a monotone decline. The invocation of the authors' Nobel standing, awarded in 2024 for work on how institutions shape national prosperity and not for this paper, lends rhetorical authority to what remains a theoretical model with restrictive assumptions. If anything, that body of work supports the institutional remedies advocated below and cautions against technological fatalism. And the description of the result as the most alarming conclusion in the AI literature mistakes a conditional theoretical possibility for a confirmed empirical prediction.

Correcting these slippages prepares the ground for taking the model seriously; it does not defend complacency, because a hypothesis stated precisely can be tested, calibrated, and acted upon, whereas a slogan can only be believed or disbelieved. The forecasting community has a particular responsibility here: its credibility depends on resisting the compression of conditional models into unconditional headlines, even, and especially, when the headline is congenial to a prevailing mood of technological apprehension.

\section{What is durable, and a forward agenda}

Stripped of its more catastrophic framing, the model identifies something real and important: a structural negative externality that standard accounts of AI overlook. When learning is a public good produced as a by-product of private effort, and when AI reduces the incentive to expend that effort, the market alone cannot sustain the optimal level of general knowledge. This is a tragedy of the cognitive commons, structurally analogous to the classical commons problems analyzed by Hardin (1968) and Ostrom (1990), and to the specific treatment of knowledge as a commons developed by Hess and Ostrom (2007). 

Recent work by Lovett (2026) examines this same structural risk through the lens of human resource development, tracing how AI adoption disrupts the regeneration of professional expertise and introducing the concepts of Internalized Mastery and Distributed Mastery to capture the distinction between deep domain knowledge and fluency in orchestrating human-AI systems. While Lovett provides the descriptive architecture and institutional vocabulary for understanding how organizations deplete shared expertise, our analysis supplies the formal economic mechanism and the calibration agenda that determine whether a collapse region exists. The two approaches are complementary.

It rests, too, on an older insight: knowledge spillovers, precisely because they are imperfectly appropriable, are under-supplied by the market even as they drive long-run growth (Romer, 1990). The contribution is to show that agentic AI sharpens this externality by severing the link between private benefit and public contribution. The concept is not wholly novel, the problem of knowledge collapse having been named in the prior literature (Peterson, 2025), but the formalization of the effort-mediated feedback loop is a genuine advance.

The most robust implication is the positive policy corollary that follows from the model. Whatever the realized value of effort elasticity, strengthening the aggregation of validated human knowledge is unambiguously beneficial. Communities of practice, professional standards bodies, open repositories, and universities acting as stewards of validated knowledge all serve to thicken the public signal and to internalize the learning externality. Unlike the proposal to degrade AI precision, this intervention requires no developer to sabotage its product and no fragile international accord; it operates on the human side of the system and is robust to a wide range of assumptions about the machine side. It is, in the language of the model, a way of increasing the productivity of public-signal generation and limiting how much private effort is displaced.

\subsection{A research agenda}

This reframing suggests a concrete research and forecasting agenda. First, and most urgently, effort elasticity must be measured. Field and quasi-experimental designs (Campbell \& Stanley, 1963) across occupations could estimate how sharply human effort responds to the introduction of capable AI assistants, exploiting staggered rollouts in the manner of recent workplace studies of generative AI (Brynjolfsson et al., 2025; Noy \& Zhang, 2023). The distribution of these elasticities across domains would determine whether and where the economy sits in the collapse region, because the outcome depends on cross-domain variation and cannot be captured by a single average value. Cross-national variation offers a natural source of contrast, since systems that bind effort tightly to licensure and formal training should display lower elasticity than those that do not, which makes the parameter relevant to geopolitical as well as sectoral forecasting.

Second, indicators of public-good erosion should be monitored systematically. The Stack Overflow decline (del Rio-Chanona et al., 2024) demonstrates that the thinning of public knowledge sharing is measurable in real time where the commons is digital and instrumented. Analogous indicators could be constructed for other domains: contribution rates to open repositories, the diversity and novelty of publicly shared problem-solving, and the rate at which new general knowledge enters shared stocks. A dashboard of cognitive-commons health would convert an abstract externality into a governable quantity. Such indicators would also link the macro externality to the micro finding that human groups exhibit a measurable collective-intelligence factor (Woolley et al., 2010), the group-level analogue of the very stock the model places at risk. Crucially, these gauges must track persistence over time, since the welfare result depends on a sustained thinning of the commons and transient adjustments associated with new tools do not determine the long-run outcome.

Third, the endogenous contribution of AI to the general stock should be incorporated into the model as an explicit component of knowledge dynamics. As scientific foundation models mature, the relevant question becomes how to design AI and its surrounding institutions so that machine-generated insights are validated and absorbed into the public stock, turning AI from a pure consumer of the commons into a net contributor. This is a design problem, and design problems are tractable in a way that civilizational prophecies are not.

Fourth, the design of endogenous compensation mechanisms deserves study alongside formal institutions. Reputation systems that reward public contribution, as on open-source and preprint platforms, business models that condition free access to AI on the contribution of data or corrections, and proof-of-contribution schemes that credit validated additions to shared knowledge could internalize the externality through incentives, challenging the binary between market failure and public intervention that frames much of the debate.

Fifth, governance research should concentrate on the feasible horn of the policy recommendation. The aggregation institutions that counteract the externality are largely national or sectoral and do not require the contested international coordination that throttling precision would demand. Designing them well means importing the core principles of commons governance: clear boundaries around what counts as a validated contribution, graduated recognition for contributors, and polycentric forms of stewardship. The commons should serve as an operational framework, not merely a metaphor (Ostrom, 1990). Understanding how to fund and sustain such institutions, and how universities in particular can act as stewards of validated knowledge in an age of cheap generation, is a more productive line of inquiry than the pursuit of globally coordinated precision caps. The geopolitical asymmetry noted earlier reinforces the point: states that lead in capability will resist even benign coordination on precision, so the decisive advantage of aggregation institutions is that many can be built unilaterally, at national or sectoral scale, without waiting for an accord that the structure of informational rents makes unlikely (Fetzer, 2026).

Two theoretical questions underpin this agenda. The first concerns the functional form of the complementarity $Q(G,s)$: whether it is multiplicative, of constant-elasticity-of-substitution type, or marked by sharp non-linear thresholds determines whether collapse arrives gradually or abruptly, and is in principle recoverable from decision-quality data. The second concerns whether aggregation institutions and precision regulation are substitutes or complements, since the answer governs how a policymaker should combine them. Both questions are falsifiable, and answering them would move the debate from rhetoric toward calibration.

\subsection{Implications for forecasting and scenario analysis}

For the technological-forecasting community specifically, the model carries a methodological lesson that outlasts any verdict on its central claim. Standard forecasts of AI's economic impact aggregate task-level productivity effects into a macroeconomic projection, an approach exemplified by the simple-macroeconomics framework that bounds aggregate gains through task shares and cost savings (Acemoglu, 2024). Such frameworks are, by construction, blind to stock effects on collective knowledge, because they model AI as a point-in-time change in the production function and overlook its effects on the slowly evolving substrate of general understanding. A forecast that captures only the flow of productivity gains while ignoring the stock of cognitive capital will be biased optimistic precisely when the dynamic externality is strongest, and the bias will be invisible in the short run, emerging only as the commons thins over a horizon of years. The knowledge-collapse model is best read, in this light, as a correction to the scope of conventional forecasts, extending attention to stock effects on collective knowledge.

Operationalizing this correction requires scenario analysis alongside point forecasting. Because the outcome bifurcates on the value of effort elasticity, the responsible forecast is not a single trajectory but a branching set of three. In a cognitive renaissance, redeployed effort and AI-generated general knowledge (a positive $\gamma$ in equation (5)) sustain or thicken the commons, so collective intelligence rises. In a cognitive stagnation, private productivity climbs while the commons thins gradually without collapsing. In a cognitive collapse, confined to specific high-elasticity domains, the self-undermining loop dominates and both the commons and the reliability of the systems that draw on it decline. Assigning provisional probabilities to these branches requires exactly the elasticity estimates proposed in Section 5.3, which is why measurement is logically prior to forecasting here. Until those estimates exist, the appropriate forecasting posture is explicit conditionality, reporting the branch points and the parameters that govern them while keeping the uncertainty visible and avoiding its compression into a headline number. This is also the posture that best inoculates the field against the viral compression documented in Section 6, since a forecast stated as a conditional branch cannot be honestly reduced to a slogan.

These outcomes need not be shared across countries. Because effort elasticity is shaped by nationally specific institutions of training, licensure, and professional identity, one economy can drift toward collapse in a domain while another sustains its commons, which raises the question of whether cognitive commons are national or global goods. General knowledge that is freely shared diffuses across borders, so a country that under-invests in its own commons may free-ride on others; yet the tacit and context-specific layers that complement it are far less mobile. The salient risk for sovereignty is therefore asymmetric: a nation can import general knowledge while losing the domestic capacity to generate and critically absorb it, a dependency that the geopolitics of informational rents (Fetzer, 2026) suggests will deepen existing divides in AI capability.

A final implication concerns leading indicators. The value of the Stack Overflow finding (del Rio-Chanona et al., 2024) is that it demonstrates the feasibility of a real-time gauge of commons health in at least one domain. A forecasting program could assemble a portfolio of such gauges across instrumented domains, treating sustained divergence between private productivity and public contribution as an early warning that an economy is drifting toward the stagnation or collapse branch. Indicators of this kind would give policymakers the lead time that the slow dynamics of knowledge stocks would otherwise deny them, converting a structural risk that is easy to recognize only in retrospect into one that can be monitored prospectively.

\section{Conclusion}

The knowledge-collapse model is neither the proof of impending catastrophe that its viral image suggests nor the alarmist overreach that its critics sometimes allege. It is a serious and partly evidenced articulation of a real risk: that agentic AI, by substituting for human effort without replenishing the collective knowledge that effort produces, can degrade the shared substrate on which both human judgment and machine reliability ultimately depend. Its central result is conditional, hinging on an effort elasticity the paper does not measure, and it rests on assumptions, a frozen knowledge taxonomy and an AI that cannot create general knowledge, that are increasingly difficult to sustain. Its proposed remedy of deliberately imprecise AI is politically and competitively infeasible, with the exceptions identified in Section 4.5.

What survives scrutiny is the diagnosis of an externality and the constructive response it implies. The tragedy of the cognitive commons is a coherent and consequential idea, and the unambiguous policy lesson is to invest in the institutions that aggregate and validate human knowledge. For the forecasting community, the model's value lies less in its most dramatic conclusion than in the channel it makes visible. The long-run cost of cognitive automation may emerge through a thinning commons, a dynamic that assessments focused only on present productivity gains are unlikely to capture. Measuring effort elasticity, monitoring the health of the cognitive commons, and strengthening the institutions that sustain it are the tasks that follow. The collapse is not guaranteed. Whether it occurs is, to a significant degree, a matter of institutional choice, which is precisely why the model is worth taking seriously and precisely why its fatalistic popular reading should be resisted.

\section*{References}
\setlength{\parindent}{-0.3in}
\setlength{\leftskip}{0.3in}
\setlength{\parskip}{8pt}
\noindent
Acemoglu, D. (2024). The simple macroeconomics of AI (NBER Working Paper No.~32487). National Bureau of Economic Research. \url{https://doi.org/10.3386/w32487}

Acemoglu, D., Kong, D., \& Ozdaglar, A. (2026). AI, human cognition and knowledge collapse (NBER Working Paper No.~34910). National Bureau of Economic Research. \url{https://doi.org/10.3386/w34910}

Acemoglu, D., Lin, T., Ozdaglar, A., \& Siderius, J. (2026). How AI aggregation affects knowledge (NBER Working Paper No.~35036). National Bureau of Economic Research. \url{https://doi.org/10.3386/w35036}

Acemoglu, D., \& Restrepo, P. (2019). Automation and new tasks: How technology displaces and reinstates labor. Journal of Economic Perspectives, 33(2), 3-30. \url{https://doi.org/10.1257/jep.33.2.3}

Brynjolfsson, E., Li, D., \& Raymond, L. (2025). Generative AI at work. The Quarterly Journal of Economics, 140(2), 889-942. \url{https://doi.org/10.1093/qje/qjae044}

Callaway, B., \& Sant'Anna, P. H. C. (2021). Difference-in-differences with multiple time periods. Journal of Econometrics, 225(2), 200-230. \url{https://doi.org/10.1016/j.jeconom.2020.12.001}

Campbell, D. T., \& Stanley, J. C. (1963). Experimental and quasi-experimental designs for research. Houghton Mifflin.

Clark, A., \& Chalmers, D. (1998). The extended mind. Analysis, 58(1), 7-19. \url{https://doi.org/10.1093/analys/58.1.7}

Cordasco, C. L. (2026, February). Acemoglu et al.~(2026) are wrong about AI and human cognition [Commentary]. \url{https://carlolc.substack.com/p/acemoglu-et-al-2026-are-wrong-about}

de Chaisemartin, C., \& D'Haultfoeuille, X. (2020). Two-way fixed effects estimators with heterogeneous treatment effects. American Economic Review, 110(9), 2964-2996. \url{https://doi.org/10.1257/aer.20181169}

del Rio-Chanona, R., Laurentsyeva, N., \& Wachs, J. (2024). Large language models reduce public knowledge sharing on online Q\&A platforms. PNAS Nexus, 3(9), pgae400. \url{https://doi.org/10.1093/pnasnexus/pgae400}

El Louadi, M. (2024). Organizational knowledge heterogeneity: Between conceptual nuances and the quest for a measure. Knowledge \& Process Management, 31(1), 17-32. \url{https://doi.org/10.1002/kpm.1756}

Fetzer, T. (2026). Discussion of ``AI, human cognition and knowledge collapse'' and the geopolitical backdrop [Commentary]. \url{https://www.trfetzer.com/discussion-of-ai-human-cognition-and-knowledge-collapse-and-the-geopolitical-backdrop/}

Galiani, S. (2026, February). When AI starts consuming the knowledge that made it valuable [Commentary]. \url{https://sebastiangaliani.substack.com/p/from-the-nber-1-when-ai-starts-consuming}

Hardin, G. (1968). The tragedy of the commons. Science, 162(3859), 1243-1248. \url{https://doi.org/10.1126/science.162.3859.1243}

Hess, C., \& Ostrom, E. (Eds.). (2007). Understanding knowledge as a commons: From theory to practice. MIT Press. \url{https://doi.org/10.7551/mitpress/6980.001.0001}

Hutchins, E. (1995). Cognition in the wild. MIT Press.

Jumper, J., Evans, R., Pritzel, A., Green, T., Figurnov, M., Ronneberger, O., Tunyasuvunakool, K., Bates, R., \v{Z}\'idek, A., Potapenko, A., Bridgland, A., Meyer, C., Kohl, S. A. A., Ballard, A. J., Cowie, A., Romera-Paredes, B., Nikolov, S., Jain, R., Adler, J., \dots Hassabis, D. (2021). Highly accurate protein structure prediction with AlphaFold. Nature, 596(7873), 583-589. \url{https://doi.org/10.1038/s41586-021-03819-2}

Kosmyna, N., Hauptmann, E., Yuan, Y. T., Situ, J., Liao, X.-H., Beresnitzky, A. V., Braunstein, I., \& Maes, P. (2025). Your brain on ChatGPT: Accumulation of cognitive debt when using an AI assistant for essay writing task (arXiv:2506.08872). arXiv. \url{https://doi.org/10.48550/arXiv.2506.08872}

Lovett, N. (2026). The tragedy of the cognitive commons: How AI could disrupt the regeneration of professional expertise. Human Resource Development Review. Advance online publication. \url{https://doi.org/10.1177/15344843261470602}

Noy, S., \& Zhang, W. (2023). Experimental evidence on the productivity effects of generative artificial intelligence. Science, 381(6654), 187-192. \url{https://doi.org/10.1126/science.adh2586}

Ostrom, E. (1990). Governing the commons: The evolution of institutions for collective action. Cambridge University Press. \url{https://doi.org/10.1017/CBO9780511807763}

Peterson, A. J. (2025). AI and the problem of knowledge collapse. AI \& Society, 40(5), 3249-3269. \url{https://doi.org/10.1007/s00146-024-02173-x}

Plato. (1995). Phaedrus (A. Nehamas \& P. Woodruff, Trans.). Hackett Publishing. (Original work composed ca. 370 BCE)

Romer, P. M. (1990). Endogenous technological change. Journal of Political Economy, 98(5, Part 2), S71-S102. \url{https://doi.org/10.1086/261725}

Shumailov, I., Shumaylov, Z., Zhao, Y., Papernot, N., Anderson, R., \& Gal, Y. (2024). AI models collapse when trained on recursively generated data. Nature, 631(8022), 755-759. \url{https://doi.org/10.1038/s41586-024-07566-y}

Woolley, A. W., Chabris, C. F., Pentland, A., Hashmi, N., \& Malone, T. W. (2010). Evidence for a collective intelligence factor in the performance of human groups. Science, 330(6004), 686-688. \url{https://doi.org/10.1126/science.1193147}

\end{document}